# Ferroelectric nematic liquid crystal thermo-motor

Marcell Tibor Máthé[1], Ágnes Buka[1], Antal Jákli[1,2,3], Péter Salamon[1,*]

[1]Institute for Solid State Physics and Optics, Wigner Research Centre for Physics, P.O. Box 49, Budapest H-1525, Hungary

[2]Materials Sciences Graduate Program and Advanced Materials and Liquid Crystal Institute, Kent State University, Kent, Ohio 44242, USA

[3]Department of Physics, Kent State University, Kent, Ohio 44242, USA

*: Author for correspondence: salamon.peter@wigner.hu

## Abstract

A thermal gradient-induced circular motion of particles placed on ferroelectric nematic liquid crystal sessile drops is demonstrated and explained. Unlike hurricanes and tornadoes that are the prime examples for thermal motors and where turbulent flows are apparent, here the texture without tracer particles appears completely steady indicating laminar flow. We provide a simple model showing that the tangential arrangement of the ferroelectric polarization combined with the vertical thermal gradient and the pyroelectricity of the fluid drives the rotation of the tracer particles that become electrically charged in the fluid. These observations provide a fascinating example of the unique nature of fluid ferroelectric liquid crystals.

## Introduction

In solid ferroelectrics, large pyroelectric coefficients and bound charge accumulations could be observed as a result of gradients of composition [1–4], strain [5] or temperature [6,7]. Dipole vortices were found in nanoscale solid state systems of ferromagnetic and ferroelectric materials that promise new applications as nano-memories, sensors and transducers [8]. Vortex-like topological structures of dipoles that exhibit toroidal moments [9–12], were predicted and found in nanoparticles [13–16], nanocomposites [17], thin films [18,19] of ferroelectrics and also in a metamaterial [20].

Solid ferroelectrics do not have mobile ionic charges, therefore it is an interesting question, whether fluid ferroelectrics with mobile ionic contaminations can produce similar phenomena. 3D fluid ferroelectric nematic ($N_F$) materials have been observed recently in liquid crystals (LCs). The $N_F$ LC phase is the newest form of nematic ($N$) LCs that are uniaxial anisotropic fluids which can be reversibly reoriented with electric fields, making nematics the essential elements of the dominant technology for electronic information devices, such as today's flat panel displays. While in normal dielectric nematic liquid crystals the average molecular axis, the director $\hat{n}$, has a head-tail symmetry, i.e., $\hat{n} = -\hat{n}$, in ferroelectric nematic liquid crystals the director is a vector, $\vec{n}$, which is parallel to the average molecular dipole density, the macroscopic polarization, $\vec{P_s}$. Although it was predicted by Born already in 1916 [21,22], there were no unambiguous experimental indications of a fluid ferroelectric nematic phase until the syntheses of the highly polar rod-shaped compounds referred respectively as DIO and RM734 by Nishikawa et al. [23] and Mandle et al. [24,25] in 2017. These materials have a large dipole moment of about 10 Debye, a ferroelectric polarization up to 5 µC/cm² and as high as ε ~ $10^4$ dielectric permittivity. The $N_F$ phase of RM734 was first suggested to have splayed polar order [26–28], but more recently it was shown that it has a uniform ferroelectric nematic phase [29].

Here we show that both the appearance of the polarization gradient induced bound charges and the vortex structure observed in solid ferroelectrics can manifest themselves in form of thermal gradient-induced circular motion of particles placed on ferroelectric nematic liquid crystal drops. In contrast to hurricanes, tornadoes, heat powered turbines and geothermal pumps that are prime examples for thermal motors and where turbulent motions are apparent, here the texture without tracer particles appears completely steady indicating laminar flow. We will show that the tangential arrangement of the ferroelectric polarization combined with the vertical thermal gradient and the pyroelectricity of the fluid drives the rotation of the tracer particles that become electrically charged in the fluid.

**Experimental results**

In our studies, we used the previously reported compound 4-[(4-nitrophe-noxy)carbonyl]phenyl2,4-dimethoxybenzoate (RM734) [24–27,29]. The flat glass substrates, served as base plates of the sessile droplets, were spin coated by a polyimide (JALS204, JSR, Japan) providing perpendicular orientation of director to the surface in the nematic phase. The sessile droplets were prepared in a custom-made setup with heated environment and micromanipulators. The experiments were carried out by using a Leica DMRX polarizing microscope equipped with a Linkam LTS350 hot stage and a TMS94 controller providing a temperature stability of 0.01°C.

Polarized optical microscopy images of sessile sub-millimeter diameter drops of RM734 heated from the bottom are shown in Figure 1 at four different temperatures near the $N - N_F$ transition.

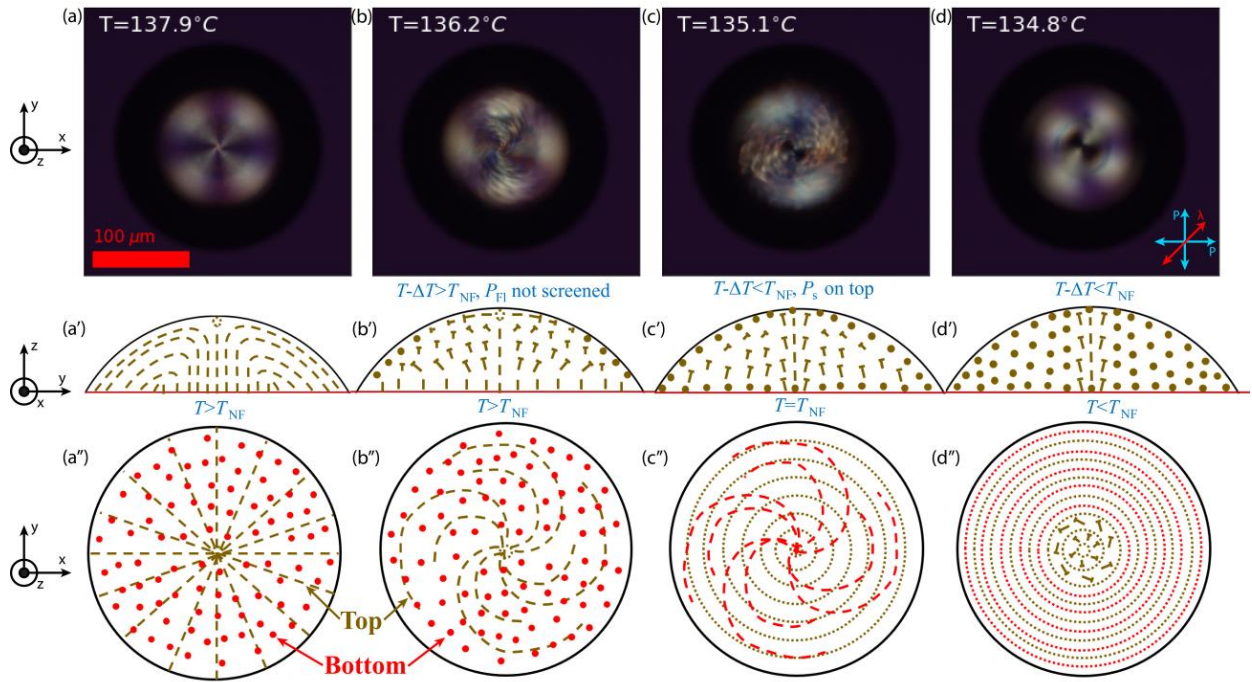

*Fig. 1: Polarizing optical micrographs and structures of the sessile drop around the $N \to N_F$ transition. Top temperature: $T - \Delta T$, bottom temperature: $T$. (a) Radial director structure in the $N$ phase ($T - \Delta T > T_{NF}$). The top temperature approaches (b), then goes below $T_{NF}$ (c). The entire drop is in the $N_F$ phase (d). The primed and double primed figures show the corresponding side and top cross sections of the director structure, respectively.*

In the $N$ phase, the director is perpendicular to the solid substrate ("homeotropic" alignment) achieved by a thin polyimide coating, and verified by independent measurements in sandwich cells. The four-brush texture seen in Figure 1(a) corresponds to a radial director structure with a central defect at the top, and a parallel (or tilted) director at the upper curved interface, as sketched in Figure 1(a', a''). For simplicity, we display parallel director there. When the colder top approaches $T_{NF}$, the splay elastic constant $K_{11}$ decreases, presumably due to the presence of a flexoelectric polarization $\vec{P}_{Fl} = e_1(\nabla \cdot \hat{n})$ induced by the splay of the radial director structure [27]. If $P_{Fl}$ is large enough, the radial flexoelectric polarization cannot be screened by free charges anymore. Consequently, an internal electric field $\vec{E}_{in} = -\frac{\vec{P}_{Fl}}{\varepsilon_o \varepsilon_\parallel}$ forms that enforces a spiralling director distribution at the top to prevent the polarization from ending in an insulating surface, leading to the texture and director structures seen in Figure 1(b, b', and b''). Further cooling through the $N \to N_F$ transition at the colder top, a ferroelectric polarization $\vec{P}_s$ is generated leading to an internal field $\vec{E}_{in} = -\frac{\vec{P}_s}{\varepsilon_o \varepsilon_\parallel}$, and the polarization field becomes tangential. Such a texture and the sketch of the corresponding director configuration are seen in Figure 1(c, c' and c''). As the ferroelectric phase reaches the bottom, the polar homeotropic surface orientation would result in an internal field that turns the director and the polarization tangentially in the entire droplet except in the vicinity of the central defect. This texture and the corresponding director structures are seen in Figure 1(d, d' and d''). Note that by using the apparent temperature difference of the phase transitions on the top (Fig.1(b)) and on the bottom (Fig.1(c)) combined with the known height of the droplet (47 µm), we can estimate the temperature gradient as 23 mK/µm.

Polarimetric measurements of RM734 in cylindrical basins of about 850 nm depth support the above scenario. A custom modification of our polarizing microscope allowed polarimetric measurement using tuneable liquid crystal retarders [30] providing two dimensional spatial maps of the average director field and the magnitude of optical retardation due to birefringence. The distribution of the average director is represented by a field of green rods in Figure 2. The initial radial configuration (Figure 2(a)) starts to be spiralling, and at the same time, the increase of retardation (Figure 2(b)) indicates that the homeotropic region at the bottom shrinks while still being in the $N$ phase. In the polar phase, the director becomes tangential in the entire drop (Figure 2(c)), except near the central defect, where the reduced retardation suggests that the director remains homeotropic in the middle.

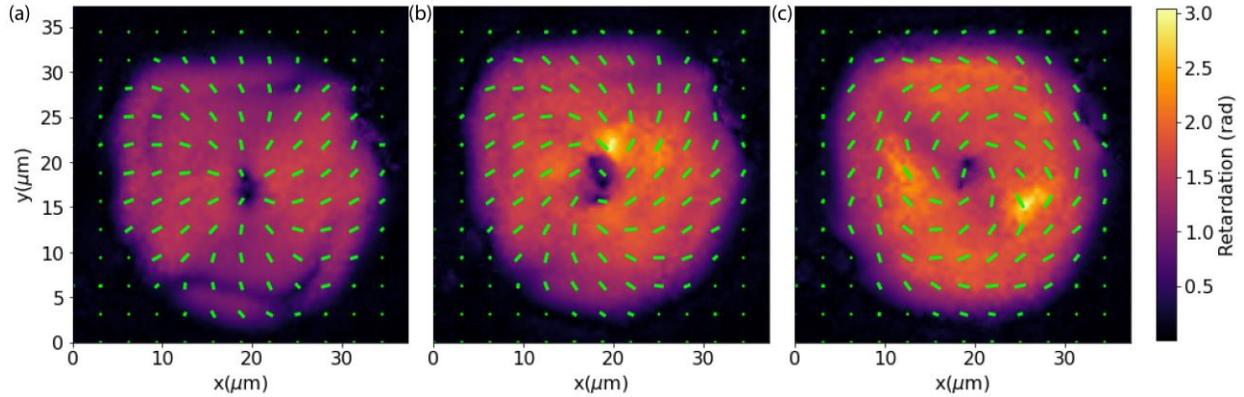

*Figure 2: The retardation of a thin disc of RM734. (a) at 140°C in nematic phase, (b) at 134°C close to the N-$N_F$ transition, (c) at 127°C in $N_F$ phase. The green rods represent the averaged orientation of the director.*

In one drop, we observed a few micrometer diameter particle, which is at rest in the *N* phase but starts circulating when it is cooled from the *N* to the $N_F$ phase. The circular flow starts immediately, as soon as the upper part of the drop went under the phase transition, as seen in Supplementary Video 1. Then 8 μm diameter buoyant polystyrene microspheres (Sigma-Aldrich) were intentionally placed on the drops as tracers. Even larger clusters of particles were able to circulate as floating on the droplets (see Supplementary Video 2). The rotation at constant 131°C in the $N_F$ phase is illustrated in Figure 3(a-d) by showing snapshots of the drops with the particles on their top. The height of the droplet in this example was 66 μm. The green and yellow arrows pointing to two clusters help to follow the clockwise motion around the central part of the droplet, where a topological defect is located.

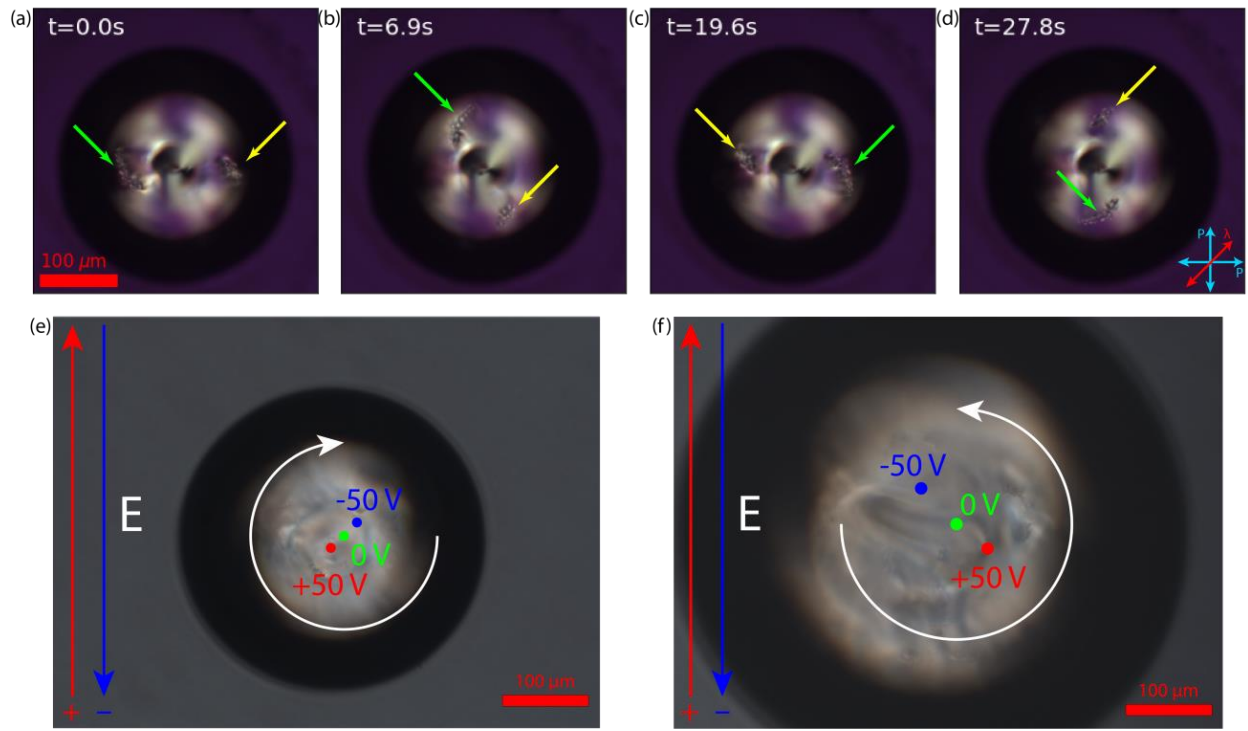

*Fig.3: Time series of snapshots of a sessile droplet of RM734 (at 131°C in the ferroelectric nematic phase) taken in a polarizing microscope with crossed polarizers and a tint-plate (a-d). The two arrows indicate two groups of tracer particles following a circular orbit around a central defect. In-plane dc electric field induced displacement of the central defect core (indicated by colourful dots) subjected to ±50 V in case of two droplets: with clockwise (e) and counter-clockwise (f) flow.*

We noticed that the presence of a vertical temperature gradient plays key role in the circular flow found in the ferroelectric droplets. The following experimental observations support this hypothesis: 1) keeping the bottom substrate at a constant temperature while cooling the top increases the speed of rotation; 2) turning the drops upside down results in opposite rotation direction; 3) increasing the temperature at the top by a transparent heater allows to completely stop and even to invert the circulation; then turning off the top heater restores the original state as seen in Supporting Video 3.

The angular velocity of the rotation is found to be proportional to the thermal gradient across the drop, while the LC director appears to be at rest. This is in contrast to the famous Lehmann rotation of chiral nematic liquid crystal drops, where the molecular orientation rotates under a thermal gradient [31,32].

The rotation direction is random from one drop to another (see Supporting Video 2), but it remains the same if the material does not undergo a phase transition. When heating up and cooling down again to the ferroelectric phase, the direction of the rotation may change. The speed of rotation is lower for smaller droplets. In some bigger droplets, two defects could be observed, and in those cases, the rotation was observable around the defect cores with opposite direction as seen in Supporting Video 4.

The circulation is found to exist as long as the temperature gradient is maintained, although we experienced a decay of the rotation speed of the particles after several hours, e.g. about 40% in 4 hours. This may be attributed to the metastability of vortex states reported in small ferroelectric capacitors of circular confinement [33] due to magnetic induction ($\vec{\nabla} \times \vec{D} = -\frac{d\vec{B}}{dt}$) and the chemical retardation of the material upon long term thermal exposure.

A further important experimental finding indicated that the positions of the upper defect cores can be affected by dc electric fields applied in the sample plane using linear surface electrodes (placed 1 mm away from each other). In the distinct droplets presented in Fig. 3(e) and Fig. 3(f), we can see two spectacular features. The first is that the defect cores are displaced in the sample plane in a direction not parallel but at an angle of about 45° with respect to the external electric field. The displacement direction depends on the sign of the field as well as on the direction of the circular flow as seen in Figure 3(e-f). Our experiments indicated negative charge accumulation in the top defects in the observed droplets. Finally, when we applied dc electric field in vertical direction across the drop, we could occasionally observe the flipping of rotation direction. Furthermore, in case of large droplets (see a 1.4 mm diameter droplet in Supporting Video 5), we observed circular motion of defect lines in the ferroelectric phase indicating stationary rotational flow without continuous director rotation, which is not due to the presence of tracer particles.

## Discussion

Cross effects, including thermomechanical coupling were predicted theoretically in the polar nematic phase, however, their specific physical mechanism and their experimental significance have not been clarified yet [34]. Here we propose the following model to describe the experimental findings presented above. The key is to consider a sharp increase of the magnitude of the spontaneous polarization $|\vec{P_s}|$ on cooling below the $N \to N_F$ phase transition [29]. According to our experiments, $\vec{P_s}$ lies along the vertical axis in the middle of the ferroelectric droplets, as being parallel to the homeotropic director in the vicinity of the central defect. Consequently, a vertical temperature gradient (assuming a colder top) results in a higher polarization that leads to an extra (let us assume positive) bound charge (volume) density at the central top, given by $\rho_v = -\vec{\nabla} \cdot \vec{P_s}$ [35]. Radially outwards from the defect, the director becomes tangential, that leads to an annular field of polarization that results in an internal electric field $\vec{E}_{in} = -\frac{\vec{P_s}}{\varepsilon_o \varepsilon_\|}$ which has a finite rotation $\nabla \times \vec{E}_{in} = -(\varepsilon_o \varepsilon_\|)^{-1} \vec{\nabla} \times \vec{P_s} \neq 0$ in the $N_F$ phase (see Fig. 4). Note that in ferroelectric nematic materials as for RM734, the ferroelectric polarization strongly increases on cooling below the $N - N_F$ transition (pyroelectricity), resulting in a rotation that is proportional to the temperature gradient. Ionic contaminants as mobile charges of the opposite sign need to compensate the central pyroelectric bound charge; therefore, a radial charge separation occurs and the region further from the defect becomes richer in positive charges. The tangential electric field can exert a force on the charges to follow an orbit around the defect. The drag force induced by the charged molecules on neutral fluid elements can result in global flow in the droplet.

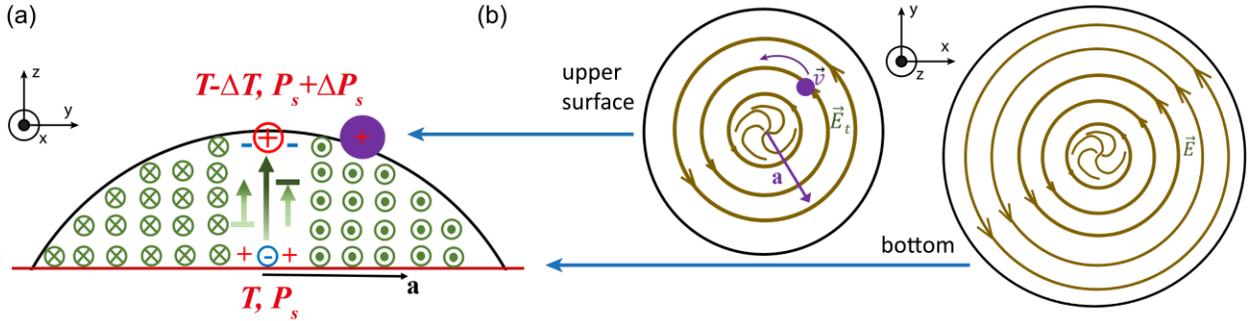

*Fig.4: Side view (a) and top view (b) sketches of a ferroelectric nematic droplet in a vertical temperature gradient leading to a tangential electric field and material flow.*

Let us consider now a simplified explanation of the tangential movement of tracer particles. Neglecting any acceleration of the particles, as a crude approximation, we can write that the electric and viscous drag forces acting on a particle orbiting at a radius $a$, are in balance: $q_p \vec{E}_t = -\vec{F}_{drag} \approx 6\pi \eta \vec{v} R$, where $q_p$, $\vec{E}_t$, $\eta$, $\vec{v}$ and $R$ are the charge of the particle, the effective electric field, the viscosity, the particle velocity, and radius, respectively. According to Poisson's equation, the depolarizing electric field is $\vec{E}_t = -\frac{\vec{P}_s}{\varepsilon_o \varepsilon_\parallel}$ [36]. The charge at the particle is proportional to the bound charge as $q_p \propto -\xi(a) \frac{\partial P_s}{\partial T} \frac{\partial T}{\partial z} V_c$, where $\xi(a)$, $p = \frac{\partial P_s}{\partial T}$, $\frac{\partial T}{\partial z}$ and $V_c$ are a quotient related to the radius dependent charge distribution, the pyroelectric coefficient, the vertical temperature gradient and the effective contact volume of the particle, respectively. Assuming that $V_c$ is proportional to the product of the contact area and a penetration depth as $V_c = R^2 A_c$, after staightforward calculations, we get an estimate on the angular frequency $\varpi$ of the particle movement as $\varpi \propto \frac{\xi(a)}{a 6\pi\eta} \frac{|\vec{P}_s|}{\varepsilon_o \varepsilon_\parallel} p \frac{\partial T}{\partial z} A_c R$. This model is obviously oversimplified, but it captures the main feature of our experimental findings, namely that the circular flow requires the presence of spontanous polarization, and the rotation speed is proportional to the temperature gradient. Considering the parameters: $R = 4~\mu m$, $A_c = 1~\mu m$, $\eta = 1~Pas$ [26], $\varepsilon_\parallel = 10^4$, $a = 80~\mu m$, $\delta z = 10~\mu m$, $T = 130~°C$, $\delta T = 1°C$, $|\vec{P}_s| = 2.7~\mu C/cm^2$ [29], $\delta|\vec{P}_s| = 0.4~\mu C/cm^2$, and $\xi = 0.5$, we get similar angular velocity ($\omega = 0.16~s^{-1}$) that found in the experiments.

The oblique displacement of defects in electric fields can be explained by the interaction of the top pyroelectric charge with the local electric field, which is the sum of the external field and the tangential field due to the vortex of spontaneous polarization. This explains the dependence of displacement direction on both the sign of the external field and the helicity of the circular flow as well. Flipping the vertical polarization at the defect using vertical dc electric field would invert the sign of the bound charge according to the model described above. In case of an unchanged tangential polarization, this would lead to the inversion of orbiting direction. In our experiments, however, we found only an occasional reversal of circulation upon applying vertical dc electric fields. We can understand this by considering that the applied field oriented the director (and the polarization) not only at the defect, but in regions of tangential orientation as well. After switching off the external field, the direction of annular polarization is randomly selected, similar to the case when cooling from the nematic to the ferroelectric phase.

## Acknowledgement

This work was financially supported by the Hungarian National Research, Development and Innovation Office under grant NKFIH FK125134 and the US National Science Foundation under grant DMR-1904167. We are thankful to Ewa Körblova and David Walba at University of Colorado at Boulder for providing RM734 for us.## References

[1]  J. V. Mantese, N. W. Schubring, A. L. Micheli, and A. B. Catalan, *Ferroelectric Thin Films with Polarization Gradients Normal to the Growth Surface*, Appl. Phys. Lett. **67**, 721 (1995).

[2]  J. V. Mantese, N. W. Schubring, A. L. Micheli, A. B. Catalan, M. S. Mohammed, R. Naik, and G. W. Auner, *Slater Model Applied to Polarization Graded Ferroelectrics*, Appl. Phys. Lett. **71**, 2047 (1997).

[3]  F. Jin, G. W. Auner, R. Naik, N. W. Schubring, J. V. Mantese, A. B. Catalan, and A. L. Micheli, *Giant Effective Pyroelectric Coefficients from Graded Ferroelectric Devices*, Appl. Phys. Lett. **73**, 2838 (1998).

[4]  M. Marvan, P. Chvosta, and J. Fousek, *Theory of Compositionally Graded Ferroelectrics and Pyroelectricity*, Appl. Phys. Lett. **86**, 221922 (2005).

[5]  Y. Zheng, B. Wang, and C. H. Woo, *Effects of Strain Gradient on Charge Offsets and Pyroelectric Properties of Ferroelectric Thin Films*, Appl. Phys. Lett. **89**, 062904 (2006).

[6]  Q. Zhang and I. Ponomareva, *Microscopic Insight into Temperature-Graded Ferroelectrics*, Phys. Rev. Lett. **105**, 147602 (2010).

[7]  Q. Zhang and I. Ponomareva, *Depolarizing Field in Temperature-Graded Ferroelectrics from an Atomistic Viewpoint*, New J. Phys. **15**, 043022 (2013).

[8]  Y. Zheng and W. J. Chen, *Characteristics and Controllability of Vortices in Ferromagnetics, Ferroelectrics, and Multiferroics*, Reports Prog. Phys. **80**, 086501 (2017).

[9]  N. Talebi, S. Guo, and P. A. Van Aken, *Theory and Applications of Toroidal Moments in Electrodynamics: Their Emergence, Characteristics, and Technological Relevance*, Nanophotonics **7**, 93 (2018).

[10]  V. M. Dubovik and V. V. Tugushev, *Toroid Moments in Electrodynamics and Solid-State Physics*, Phys. Rep. **187**, 145 (1990).

[11]  S. Prosandeev, I. Ponomareva, I. Kornev, I. Naumov, and L. Bellaiche, *Controlling Toroidal Moment by Means of an Inhomogeneous Static Field: An Ab Initio Study*, Phys.